**Gasoline Pricing Policies for Transportation Safety**

By

Nima Safaei
Graduate Student
Department of Civil, Construction, and Environmental Engineering
Iowa State University
2711 S Loop Dr.
Ames, IA 50010
Phone: (515) 715-3804
Email: nsafaei@iastate.edu

Chao Zhou
Graduate Student
Department of Civil, Construction, and Environmental Engineering
Iowa State University
2711 S Loop Dr.
Ames, IA 50010
Phone: (515) 528-5025
Email: czhou@iastate.edu, czhou115@gmail.com

**ABSTRACT**

Economic factors can have substantial effects on transportation crash trends. This study makes a comprehensive examination of the relationship between the retail gasoline price (including state and federal fuel taxes) and transportation fatal crashes from 2007 to 2016 in the US.

Data on motor vehicle, bicycle and pedestrian fatal crashes come from Fatality Analysis Reporting System (FARS) provided by the National Highway Safety Administration (NHTSA) and the gasoline price data is from U.S. Energy Information Administration (EIA). Random effect negative binomial regression models are used to estimate the impact of inflation-adjusted gasoline prices on trends of transportation fatal crashes.

Initial results combined with results of previous studies showed that gender and transportation mean type (motorcycle, non-motorcycle, bicycle and pedestrian) play prominent roles in interpreting the final model, so by using random effect negative binomial regression, seven models are developed to evaluate the effects of gasoline price changes on total population, male, female, motorcyclists, non-motorcyclists, bicyclists and pedestrians separately.

Our findings suggest that increasing the gasoline prices will not significantly alter the number of total fatal crashes. However, by looking at different vehicle types, it is estimated that one dollar increase in adjusted gasoline price is associated with 24.2% increase in the number of motorcycle fatal crashes, 1.9% decrease in the number of non-motorcycle fatal crashes, and 0.7% decrease in the number of pedestrian fatal crashes. Also, there is no noticeable difference between male and female in response to the gasoline price changes.

## INTRODUCTION

Transportation safety relies on numerous parameters ranging from pavement and infrastructure condition to the efficiency of the regulations set by the responsible authorities and their psychological implications on the road users. Recent changes in gasoline prices have led to convoluted social implications for commuters and decision makers (*1*). Higher gasoline prices have much more effects than just costing more to fill up the gas tank; it has a diverse range of effects on the broader economy and travel trends. With increased gasoline price people tend to drive less to conduct their purchases. People feel good about the economy when the gas price is lower and they tend to spend more on personal consumption such as shopping and dining (*2*). When gas price surges, people stop spending and businesses suffer from it. It has been reported that as the gasoline price increased in April 2011, online shopping in the United States reached its fastest rate in four years (*2*). Also, Marin Software reported that searches for online shopping increase significantly along with gas price (*2*). On the other hand, as the gasoline price rises, retailers need to increase the shipping cost as their vehicles need to spend more on gas. The increase in gasoline price effects the automobile industry by making automobile manufactures create small, more fuel-efficient vehicles like hybrids (*2*). Gasoline price when headed to 3.5 dollars in 2011, led to dramatic psychological effects on consumers (*3*). Rapid increase in oil price could derail the economic recovery and put into a recession *(3)*. This argument is based on the fact that 10 out of the last 11 post-World War II recessions happened immediately after oil price spikes (*3*). David Rosenberg stated that one penny increase in the gasoline price sucks 1.5 billion dollars out of annual household cash flow and drags down the GDP growth (*3*). In July 2008, gasoline price reached a record of 4 dollars per gallon and instantly people made an intense shift to buy smaller, more fuel-efficient cars (*4*). Hybrid car sales increased by 34 percent during the first season of 2011, in comparison to the first season in 2010 (*4*). Since the last season of 2014, as gasoline prices began falling, Americans moved from buying more fuel-efficient cars to larger pick-up trucks and SUVs (*4*). According to HIS, SUV and crossovers accounted for 40 percent of US market share in 2016 versus 34 percent in 2014 (*4*). It is also been noted that gasoline price ranges can have an effect on some public transportation ridership. Based on a report by the American Public Transportation Association, in April 2011, the Raleigh-Durham-Chapel Hill region of North Carolina represented an 18 percent increase in express bus riders compared to the same period in 2010 (*2*). Also, riders on New Mexico's Rail Runner increased by 14 percent during the same time (*2*). Gasoline price changes may affect the working schedule of industries. Some businesses with more flexible working schedule have adopted four-day weeks to limit their employees' financial burden for commuting during the gasoline price rising time (*2*). Also, some economists warn that gasoline price increase negatively affect the economic recovery; they have mentioned that businesses have re-evaluated their hiring strategies tending to have more lay-offs due to the uncertain economic condition. Increase in gasoline price can diminish the number of freelancers as it can limit the geographical region in which the business can flourish (*2*). Gasoline price can also affect the land and property values. According to a study conducted by Neill at the University of Nevada, based on 930,702 home sales data in the Las Vegas area for a span of more than 30 years, a 2 percent price bump is observed for homes close to the city center and a 1 percent price drop was observed for homes distant from the city center due to a 10 percent increase in gasoline prices (*2*). Gasoline price changes may also make people relocate to the areas closer to their workplaces (*5, 6*). It might be possible that as the gasoline price increases, travelers change their transportation modes and reduce their driving frequency, leading to a reduction in gasoline consumption (*1*); people might shift to using motorcycles, bicycles or public transit instead of their four-wheel vehicles; they might even choose to walk instead of riding or driving (*7*). In 2008, Mattson created a dynamic model for studying the short and long-term effects of gasoline price changes on transit-ridership use. He found that in large and large-medium urban areas, the increase in public transit-ridership in response to the gasoline price rise is considerable and it happens immediately after the change, however for smaller areas due to the less familiarity of people with public transportation systems, it took five to seven months for them to respond to the gasoline price changes and to use public transit systems instead of their personal motor vehicles (*8*). Gasoline price changes can also play an important role in transportation safety. Chi et al.'s findings suggested that if decision makers wished to reduce traffic crashes, increasing gasoline taxes would represent a possible option due to the relationship between higher gasoline prices and traffic crashes (*9*, *10*).

A comprehensive literature review was conducted on examining the effects of gasoline price changes on traffic safety. Chi et al. had studied the effects of gasoline prices on crash numbers. Using both the Poisson-gamma regression model and Prais-Winsten linear regression model, they found that when the inflation-adjusted gasoline price increased 1%, the monthly total crash number decreased 0.25% in the short term and 0.47% at a one-year lag. However, the gasoline price changes had no effects in the long term; they also found that higher gasoline prices reduced crashes for younger drivers immediately and at a one-year lag for older and male drivers. They also claimed that the gasoline price increase had both immediate and one-year-lag effects on females, whites, and blacks (*9*). In another study conducted by Chi et al. in 2013, they examined the relationship between the fluctuations in gasoline prices and traffic safety using monthly traffic crashes in the state of Mississippi from April 2004 to December 2010. They found that gasoline prices had a stronger negative correlation with less severe crashes. The effects of the gasoline price changes took place at a nine-month lag, peak at a twelve-month lag, and weaken after an eighteen-month lag. Furthermore, the results indicated that the gasoline price fluctuations had negligible effects on reducing fatal crashes. They also concluded that gasoline prices did not have an immediate effect on crash rates of male drivers. They hypothesized that as men's trips were more work-related (*11*), it would take a longer time for them to react to the new policy of gasoline prices. Women in Mississippi were responsible for more household activities, and they made more short-distance trips so they could easily change their frequent short-distance trips to less frequent multi-purpose trips (*1*). In another study conducted by Chi et al., they focused on the differences between the relationship of gasoline price and traffic crashes in urban and rural areas using the monthly traffic crash data from 1998 to 2007 at the county level in Minnesota. They found that although fatal crashes in both the urban and rural areas were strongly affected by the gasoline prices and similar decreases were observed when gasoline price increased, total crashes, property-damage-only crashes, and injury crashes were greater affected by the gasoline price changes in rural areas than urban areas (*12*). Also, in 2011, Chi et al., using the Mississippi monthly crash data from 2004 to 2008, studied the connections between gasoline prices and drunk-driving crashes. In this study, they found that higher gasoline prices would lead to fewer drunk-driving crashes. They also realized that less severe crashes were more strongly affected by gasoline price changes than fatal crashes while higher alcohol consumption may lead to more severe crashes (*13*). Chi et al., mentioning the limitations of other studies focusing either on fatal crashes only or all crashes being measured over a short period, conducted another study to examine gasoline price effects on all traffic crashes by demographic groups in the state of Alabama from 1999 to 2009 (*10*). They estimated the effects of gasoline price changes on automobile crash trends. They used age group, gender and race/ethnicity as dependent variables. They found that gasoline price changes had a stronger effect in reducing crashes involving young drivers (aged 16-20) than the older drivers in the short term. They asserted that the reasons of such an observation might be the tight budget of teenagers that made them more vulnerable to gasoline price changes, or either the rigidity of elders due to their fixed work driving routes and family responsibilities. They declared their findings to be consistent with those from some prior studies (*9*, *14*). Regarding gender, they found a contradiction between their results and their findings in the previous study (*9*). In the Alabama study (*10*), they found that there was no significant difference between male drivers and female drivers in both the short-term and long-term perspective, however, in the Mississippi study (*9*), they found that increasing gasoline prices more strongly reduced crashes involving female drivers than male drivers. Also, regarding race/ethnicity, the two studies did not come up with the same results; in the Mississippi study (*9*), they found that crashes involving non-Hispanic white drivers were more vulnerable to gasoline price changes than crashes involving non-Hispanic black drivers in the short term and the reverse in the long term. However, in the Alabama study (*10*), neither in the long term nor in the short term, there were differences between the effects of gasoline changes on crashes involving non-Hispanic white drivers and non-Hispanic black drivers. In a recent study, Chi et al., by analyzing Mississippi traffic crash data from 2004 to 2012, found that roadway safety would react to the gasoline prices in a nine to ten-month lag regardless of age, gender and races with some exceptions (*15*).

Regarding the effects of gasoline taxes, a paper was written by Grabowski et al. to understand the impact of gasoline taxes on the traffic fatality rate. They used a semi-log model on the total number of traffic fatalities for the 48 states from 1983 to 2000 with the total of 912 observations. They found that the traffic fatality rate dropped by 0.6% with a 10% increase in the gasoline tax; they also argued that the tax that led to a noticeable increase in the gasoline price would decrease traffic fatalities considerably (*14*). In another paper by Leigh et al., they found that people drove less or slower with a higher gasoline tax, which in turn reduced fatalities. They mentioned that gasoline taxes affected fatalities indirectly through decreasing speeds and fewer miles driven. The study found that, as gasoline taxes increased by 10%, the fatalities were expected to drop by 1.8 to 2.0%. However, the long-term effects may not be as evident as the results found. In addition to reducing fatalities, tax revenues could be invested in mass transit systems which in turn would increase traffic safety. The bipartisan National Surface Transportation Infrastructure Financing echoed this idea in raising the federal gasoline taxes from 18.4 cents to 28.4 cents per gallon to increase highway funding and reduce budget deficits (*16*).

There is also a considerable amount of literature that had studied the effects of gasoline price changes on the number of fatalities instead of crash numbers. Using a panel data model, Ahangari et al., incorporated other economic factors and found that with a 10% increase in the gasoline price, traffic fatalities decreased by 2.18%. According to the study, if the unemployment rate increased by 10%, the traffic fatalities would be expected to reduce by 0.65%. Additionally, the road fatality rate decreased by 41.5% with a 10% increase in the health index factor (*17*).

Zhu et al. claimed that an increase in gasoline price is accompanied by the increased costs of hospitalization of motorcycle injuries. In one study, by using the data on inpatient hospitalizations, obtained from the Nationwide Inpatient Sample between 2001 and 2010, they built panel feasible generalized least square models to examine the association of gasoline prices with hospital utilization and cost for motorcycle and non-motorcycle motor vehicle crash injuries. They found that unless the increase in the gasoline tax was integrated with public health intervention policies to improve the motorcycle safety, increased gasoline prices would cause as much rise in hospitalization cost of motorcyclists that would compensate for any reductions in the number of non-motorcycle injuries (*18*). In a prior study, Zhu et al., had chosen California as their study field as California had the highest number of motorcycle registrations in the USA, and it was ranked third in number of motorcycle crashes which accounted for the 8% of all motor vehicle fatalities and 13% of injuries from motorcycle crashes in the USA. By using the Statewide Integrated Traffic Records System for 2002–2011, autoregressive integrated moving average (ARIMA) regression models were utilized to estimate the impact of inflation-adjusted gasoline price per gallon on trends of motorcycle injuries. Their findings suggested that raising gasoline prices increased the number of motorcycle riders on the roads and, consequently, caused more motorcycle injuries. They indicated that besides the mandatory helmet use law, implementing ways to raise risk awareness of motorcyclists, investment in alternative transportation modes like public transportation and making strict licensing tests of riding skills would help to reduce the motorcycle fatal and non-fatal injuries (*19*). Hyatt et al., conducted a study to evaluate the association between increases in gasoline price for automobile occupants and motorcycle riders and motor vehicle collision-related injury and fatality rates; they reported that although the number of injuries and fatalities in motorcycle-related crashes increased with increasing gasoline price, rates of motorcycle-related injuries and fatalities per registered vehicles remained unchanged. They concluded that it was less probable that raising the gasoline prices would affect the driver's characteristic that might result in more crashes. Instead, they told that it made more drivers change their transportation modes to motorcycling and the increased number of motorcyclists on the road was the primary cause of increased motorcycle-related fatalities and injuries (*20*). Wilson et al., mentioning that raising the fuel price was a probable reason for the popularity of motorcycling in the USA, quantified the relationship between changing the fuel price and motorcycle fatalities. Their findings suggested that in response to the gasoline price increase and to reduce their fuel expenses, people aimed to change their modes of transportation to motorcycles, and although the fatality rates of automobile crashes declined as a response to the gasoline price increase, motorcycle fatality rates had risen. They mentioned the fact that they did not study changing preferences for motorcycling instead of driving and the fact that they did not incorporate motorcycle fatalities on nonpublic roads as the limitations of their study (*7*).

Litman et al. evaluated the traffic safety impacts of various transport pricing reforms, including fuel-tax increases, efficient road and parking pricing, distance-based insurance, registration fees, and public-transit fare reductions. In addition to providing other significant economic, social, and environmental benefits, the analysis indicated that such reforms could significantly improve traffic safety by reducing the traffic risk. In overall, they asserted that reforms might be the most cost-effective safety strategies and considerable traffic safety improvements could be provided by reforming pricing policies (*21*).

This paper emphasizes the relationship between the adjusted retail gasoline price and the number of fatal crashes nationwide from 2007 to 2016. Although a sufficient number of researches have been conducted for evaluating the consequences of increasing the gasoline price, most of them were focusing on a single state or during a short period. This paper studies the whole states in the USA during a 10-year-period by making separate models based on gender and vehicle type (motorcycle, non-motorcycle, bicycle and pedestrian).

## DATA DESCRIPTION
In this study, Fatality Analysis Reporting System (FARS) was used to collect state-month data on the total number of crashes involving fatal injuries for 50 states and District of Columbia from 2007 to 2016 with a total of 6121 observations. Also, the total number of fatal crashes were separated by gender and vehicle type (motorcycle, non-motorcycle, bicycle and pedestrian) to investigate the different effects of gasoline price fluctuations between male drivers, female drivers, motorcyclists, non-motorcyclists, bicyclists, pedestrians and total. The key independent variable in this study was the US monthly retail gasoline all-grades price, which was obtained from the U.S. Energy Information Administration (EIA). Gasoline price information was retrieved for five US regions (West Coast, Rocky Mountain, Midwest, Gulf Coast and East Coats) in monthly intervals for the ten-year study period (2007-2016) and then used for each state data; the information captures the overall gasoline price trends for the entire US. Gasoline prices were adjusted for inflation in 2018 dollars.

For controlling other factors that might affect the number of fatal crashes, several control variables were added to the model. Previous studies found that economic conditions (such as unemployment rate, GDP per capita, state median household income) affected people's driving behavior which in turn affected traffic crash rates (22, 23). It is also found that most of the dramatic decline in traffic fatalities that happened between 2008 and 2012 coincided with the great economic recession (24), so economic conditions play a significant role in roadway crash trends. For incorporating the effect of economic conditions into the model, the seasonally adjusted unemployment rate for each state acquired from the Bureau of Labor Statistics was added to the model. Generally, more travels increase the chance of accidents. The vehicle miles traveled (VMT) is the other significant control variable that was used in the model; since 1986, it is proven that accident occurrence increases with VMT (25). Since the monthly VMT for each state is not available, it was decided to use yearly VMT divided by 12 to yield average monthly VMT. For analyzing the bicycle and pedestrian fatal crash data, instead of VMT, the population data was gathered from the United State Census Bureau and used as a control variable in these models. Also, the population density data is used in all the models for observing its effects on the number of fatal crashes. Furthermore, maximum speed limit and differential speed limit were derived from FHWA and added to represent the state law changes that might influence the crash rate over the 10-year study period. A study conducted by NCHRP in 2013, represented that increasing the speed limit from 55 to 65 mph and from 65 to 75 mph resulted in 28% and 13% increase in the number of fatalities. Friedman et al. studied the long-term effect of speed limit increase from 1995 to 2005; they found that 12,545 deaths and 36,583 injuries in fatal crashes were caused by increased speed limit nationwide. (26, 27) Also, the differential speed limit between cars and trucks are considered as a contributing factor for causing more deaths on highways (28). As another control variable, seat belt usage rate provided by NHTSA was added into the model; although this rate is only applicable for non-motorcycle data, it could be useful for representing the state practices to some extent. Protection system use has always proven to be efficient in reducing the number of fatal crashes. In 2005, Blows et al. stated that unbelted drivers had ten times the risk of involvement in an injury crash compared to belted drivers (29). Moreover, month and year time trend were included to capture the characteristics of each season and year that might result in the changes of crash rate. Some binary indicators were created for months and years with setting "December" and "2016" for the rest of the months and years as the base values respectively. Also, Aging of the US population has caught growing attention to safe driving issue among old drivers. In a recent study, Molnar et al. showed that older drivers (60 to 70-years-old) took fewer trips with shorter periods and they are more cautious in overnight driving and have more consistent driving habits (less high deceleration and acceleration events) (30). During 1996 to 2006, although the male population was equal to or less than the female population among all age groups, motor vehicle crash fatalities were higher for male than female drivers. Additionally, it was reported that among females, the

over-65 age group had the highest number of fatalities after the 16-to-20 age group. Consequently, the percentage of senescent people (over 70 years old) and male percent were added to the control factors as they might affect the number of fatal crashes (*31*).

The data were aggregated at the monthly level in each state. Descriptive statistics for the dataset are shown in Table 1.

**TABLE 1 Descriptive Statistics of Fatal Crashes at the Monthly Level in the US, 2007-2016**

| Variable | Minimum | Maximum | Mean | Std. Dev. |
|---|---|---|---|---|
| Percent Male | 47.22 | 52.65 | 49.34 | 0.81 |
| Percent Age 70 and Above | 4.35 | 13.6 | 9.47 | 1.37 |
| Seat belt Usage Rate | 0.64 | 0.98 | 0.86 | 0.07 |
| Monthly Total Fatal Crash | 0 | 338 | 45.77 | 47.68 |
| Monthly Male Fatal Crash | 0 | 273 | 33.79 | 35.87 |
| Monthly Female Fatal Crash | 0 | 114 | 13.98 | 14.4 |
| Monthly Motorcycle Fatal Crash | 0 | 76 | 7.73 | 10.48 |
| Monthly Non-motorcycle Fatal Crash | 0 | 227 | 35.56 | 34.81 |
| Monthly Pedestrian Fatal Crash | 0 | 118 | 7.73 | 11.6 |
| Monthly Bicyclist Fatal Crash | 0 | 20 | 1.18 | 2.34 |
| Monthly Vehicle Miles Travelled (millions) | 293.92 | 28342.92 | 4925.83 | 5082.48 |
| Adjusted Retail Gasoline Price (2018 Dollars) | 0 | 5.15 | 3.38 | 0.7 |
| Seasonally Adjusted Unemployment Rate (%) | 2.4 | 14.6 | 6.48 | 2.21 |
| 65-mph Speed Limit and below (0-No, 1-Yes) | 0 | 1 | 0.31 | 0.46 |
| 70-mph Speed Limit (0-No, 1-Yes) | 0 | 1 | 0.4 | 0.49 |
| 75-mph Speed Limit and Above (0-No, 1-Yes) | 0 | 1 | 0.29 | 0.45 |
| Differential Speed Limit (0-No, 1-Yes) | 0 | 1 | 0.17 | 0.37 |
| West Coast (0-No, 1-Yes) | 0 | 1 | 0.14 | 0.34 |
| Rocky Mountain (0-No, 1-Yes) | 0 | 1 | 0.1 | 0.3 |
| Midwest (0-No, 1-Yes) | 0 | 1 | 0.29 | 0.46 |
| Gulf Coast (0-No, 1-Yes) | 0 | 1 | 0.12 | 0.32 |
| East Coast (0-No, 1-Yes) | 0 | 1 | 0.35 | 0.48 |
| Population | 534876 | 39250017 | 6128032 | 6874432 |
| Population Density | 1.192 | 11157.58 | 393.07 | 1421.50 |
| January | 0 | 1 | 0.07 | 0.25 |
| February | 0 | 1 | 0.07 | 0.25 |
| March | 0 | 1 | 0.07 | 0.25 |
| April | 0 | 1 | 0.07 | 0.25 |
| May | 0 | 1 | 0.07 | 0.25 |
| June | 0 | 1 | 0.07 | 0.25 |
| July | 0 | 1 | 0.07 | 0.25 |
| August | 0 | 1 | 0.07 | 0.25 |
| September | 0 | 1 | 0.07 | 0.25 |
| October | 0 | 1 | 0.07 | 0.25 |
| November | 0 | 1 | 0.07 | 0.25 |
| December | 0 | 1 | 0.07 | 0.25 |
| 2007 | 0 | 1 | 0.1 | 0.3 |
| 2008 | 0 | 1 | 0.1 | 0.3 |
| 2009 | 0 | 1 | 0.1 | 0.3 |
| 2010 | 0 | 1 | 0.1 | 0.3 |
| 2011 | 0 | 1 | 0.1 | 0.3 |
| 2012 | 0 | 1 | 0.1 | 0.3 |
| 2013 | 0 | 1 | 0.1 | 0.3 |
| 2014 | 0 | 1 | 0.1 | 0.3 |
| 2015 | 0 | 1 | 0.1 | 0.3 |
| 2016 | 0 | 1 | 0.1 | 0.3 |

Figure 1 shows the adjusted retail gasoline price trend. The fluctuation of gas price is consistent across regions in the US, and the West Coast has a higher gasoline price typically compared to other regions in the study period. It can be observed that the retail gasoline prices have experienced two steep drops since 2007. The drop in 2008 was primarily due to the recession and the other one in 2015 was a combined result of demand and policy changes.

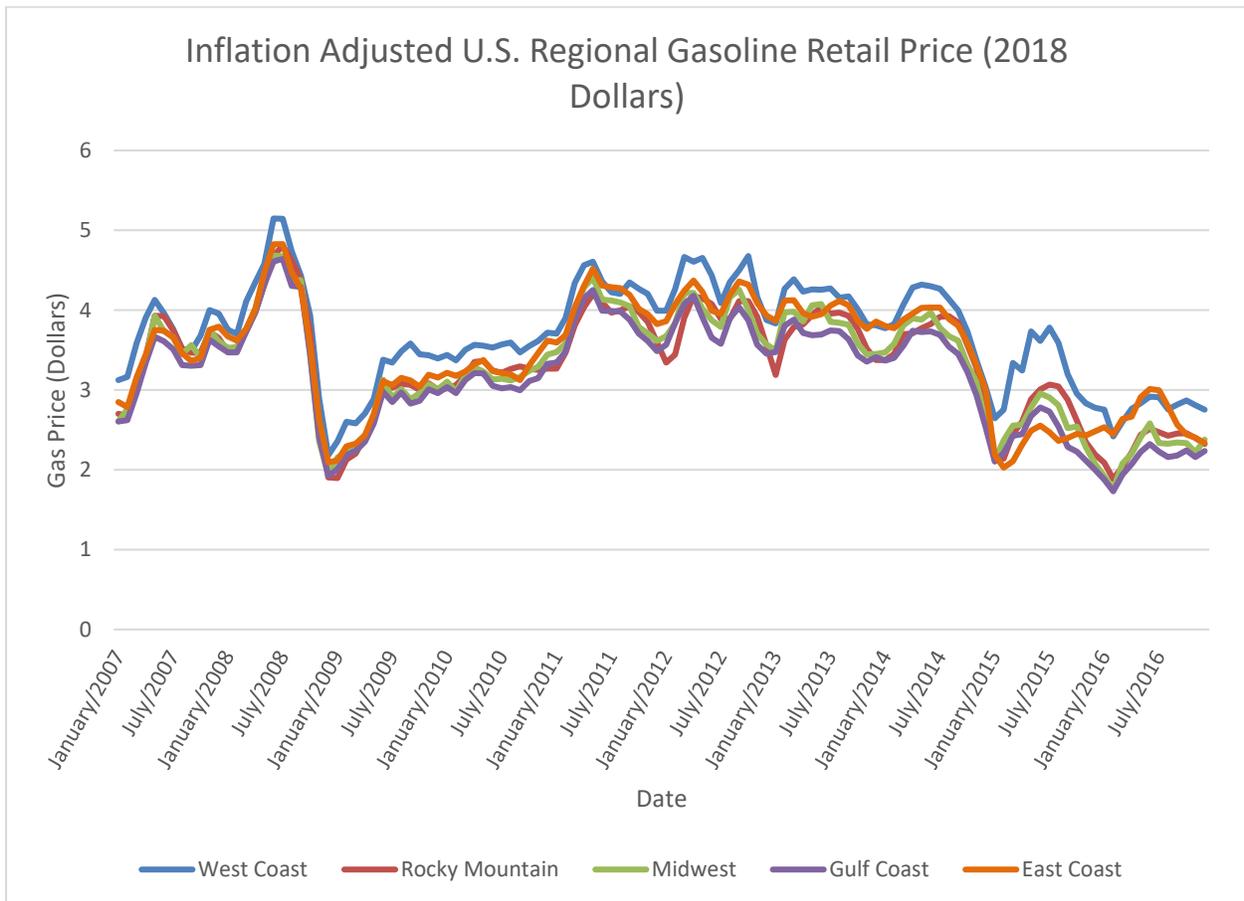

**FIGURE 1 Inflation Adjusted Monthly Gasoline Retail Price Trend, 2007-2016.**

## STATISTICAL METHODOLOGY

The study uses fatal crash data that are comprised of non-negative integers. When dealing with such count data, Poisson and negative binomial are the two most commonly used models.

Starting with the Poisson model, the probability of the number of fatal crashes equals $y_i$ at specific states during a one-month period which is shown in Equation 1.

$$P(Y = y_i) = \frac{EXP(-\lambda_i)\lambda_i^{y_i}}{y_i!}, y_i = 0,1,2,\ldots \tag{1}$$

Where $\lambda_i$ is the mean or expected value of a Poisson distribution, which in this case stands for the expected number of fatal crashes that could occur statewide at a given month. The expected number of fatal crashes is given by Equation 2 to introduce the set of explanatory variables:

$$\lambda_i = exp\ (\beta X_i)\ \text{Or}\ Ln\lambda_i = \beta X_i \tag{2}$$

Where $X_i$ is the explanatory variable and $\beta$ is the estimated parameter.

The limitation of the Poisson model is that it assumes that the variance is equal to mean, which is often not true in real data. The assumption of the Poisson model makes it unable to address overdispersion. As our analysis suggests that the fatal crash data are overdispersed, a negative binomial model is preferred over the Poisson model. Negative binomial models handle overdispersion by adding an unobserved heterogeneity term $u_i$ to the log linear as shown in Equation 3:

$$Lny_i = Ln\lambda_i + Lnu_i = \beta X_i + \varepsilon_i \tag{3}$$

Thus, the probability of the number of fatal crashes, $y_i$, which occurs at specific states during a one-month period can be re-written in Equation 4:

$$P(Y = y_i) = \frac{EXP(-\lambda_i u_i)(\lambda_i u_i)^{y_i}}{y_i!}, y_i = 0,1,2,\ldots \tag{4}$$

Unlike the Poisson model, the negative binomial model adds parameter $\alpha$ in the formula that describes the relationship between variance and mean, which can be expressed in Equation 5:

$$VAR(y_i) = E[y_i]\{1 + \alpha E(Y_i)\} \tag{5}$$

From this equation, it is observed that when $\alpha$ is equal to zero, the negative binomial model is transformed to the Poisson model.

For achieving good results with the Poisson and the negative binomial models, the crash data should be uncorrelated in time. In this study, both models seem to be inappropriate as unobserved heterogeneity and serial correlation are present in the crash data. Random effect negative binomial (RENB) model is a more suitable alternative. It can deal with the spatial and temporal effects in the data set by treating the data in a time-series cross-section panel (*32*). The

RENB model is expressed in Equation 6:

$$E(A_{it}) = exp(\beta X_{it} + \mu_i + \varepsilon_{it})$$ (6)

Where $E(A_{it})$ stands for the predicted number of monthly fatal crashes in state i in month t, $X_{it}$ is a vector of explanatory variables, $\beta$ is a vector of estimable parameters, $\varepsilon_{it}$ is the vector of residual errors, and $\mu_i$ is the random effects for the $i^{th}$ state.

As for the interpretation of coefficients, if $X_i$ is continuous, the percent change in mean response when $X_i$ is increased by one unit and the other X variables are held constant, is given in Equation 7:

$$100 \times \left[ exp(\hat{\beta}_1) - 1 \right]$$ (7)

If $X_i$ is binary, the percent change in mean response when $X_i$ is equal to one, and other X variables are held fixed can also be expressed as Equation 7.

## RESULTS AND DISCUSSION

From 2007 to 2011, the yearly total number of fatal crashes had decreased consistently from 37,435 to 24,651 nationally, 34% decrease since 2007, a period in which inflation-adjusted national average gasoline prices increased about $0.53 per gallon. By analyzing the raw data, it is calculated that the total number of fatal crashes nationwide has decreased by 9,903 in 2016 since 2007.

Random effect negative binomial models are estimated to examine the factors that may affect crash rates with a principal interest in retail gasoline prices. Other factors like seat belt use rate, natural log of vehicle miles travelled, maximum speed limit over the ten years study period for each state, the implementation of differential speed limit in the state, seasonally adjusted unemployment rate, percent male and percent older people in each state, natural log of population and population density, indicator variable for districts, month time trend, and year time trend were set as control variables. Seven models are developed: one for monthly total crash numbers, the two others for monthly male and female crash numbers, and four for different vehicle types (motorcycle, non-motorcycle, bicycle and pedestrian). The results are presented in Table 2.

**TABLE 2 Results of Random Effect Negative Binomial Regression Models for Fatal Crash Numbers (Total, Male, and Female) at the Monthly Level in the US, 2007-2016**

| Random Effects: Group Name, State (Intercept), Number of obs: 6120, groups:State:51 | Total | | Male | | Female | |
|---|---|---|---|---|---|---|
| | Variance | Std. Dev. | Variance | Std. Dev. | Variance | Std. Dev. |
| | 0.0582 | 0.2413 | 0.05501 | 0.2345 | 0.04162 | 0.204 |
| Fixed Effects: | Estimate | Pr(>\|z\|) | Estimate | Pr(>\|z\|) | Estimate | Pr(>\|z\|) |
| (Intercept) | -9.933615 | 0.000183 | -8.5108925 | 0.00125 | -4.0707869 | 0.228968 |
| Ln(VMT) | 0.902913 | < 2e-16 | 0.9473462 | < 2e-16 | 1.0018249 | < 2e-16 |
| Adjusted Retail Gasoline Price (2018 Dollars) | 0.007552 | 0.360312 | 0.0048837 | 0.59973 | 0.0004744 | 0.967899 |
| 65-mph Speed Limit and below | 0.000668 | 0.981166 | 0.0375979 | 0.23632 | -0.0723754 | 0.077074 |

| | | | | | |
|---|---|---|---|---|---|
| 70-mph Speed Limit | 0.017801 | 0.4701 | 0.0436652 | 0.11258 | -0.0137623 | 0.701204 |
| 75-mph Speed Limit and Above (Base) | - | - | - | - | - | - |
| Differential Speed Limit | -0.052026 | 0.014284 | -0.04432 | 0.05572 | -0.0776996 | 0.002806 |
| Seatbelt Usage Rate | 0.044355 | 0.694572 | -0.0140741 | 0.91117 | 0.0183818 | 0.908293 |
| West Coast | -0.538994 | 0.000685 | -0.4698823 | 0.00189 | -0.3508471 | 0.007101 |
| East Coast | -0.411079 | 0.00506 | -0.3550874 | 0.01006 | -0.2633237 | 0.017711 |
| Rocky Mountain | -0.500073 | 0.003809 | -0.4482634 | 0.00625 | -0.2823544 | 0.035276 |
| Midwest | -0.371361 | 0.005416 | -0.3421082 | 0.00683 | -0.2633395 | 0.010562 |
| Gulf Coast (Base) | - | - | - | - | - | - |
| Seasonally Adjusted Unemployment Rate (%) | -0.039424 | < 2e-16 | -0.0431455 | < 2e-16 | -0.0302585 | 7.81E-10 |
| Percent of 70-years-old and above | 0.010171 | 0.569101 | 0.0019252 | 0.91796 | 0.0309289 | 0.118602 |
| Male Percentage | 0.131555 | 0.008658 | - | - | - | - |
| Ln (Population Density) | -0.028398 | 0.490866 | -0.0568349 | 0.14375 | -0.1394177 | 0.000164 |
| January | -0.100398 | 1.99E-11 | -0.0886287 | 2.01E-07 | -0.1217316 | 9.30E-09 |
| February | -0.190607 | < 2e-16 | -0.1707604 | < 2e-16 | -0.2252916 | < 2e-16 |
| March | -0.040166 | 7.60E-03 | 0.0005111 | 9.76E-01 | -0.1215003 | 1.51E-08 |
| April | 0.02124 | 0.161957 | 0.0704661 | 0.0000416 | -0.0842271 | 0.000101 |
| May | 0.128933 | < 2e-16 | 1.94E-01 | < 2e-16 | -0.0215065 | 3.26E-01 |
| June | 0.15784 | < 2e-16 | 0.2153096 | < 2e-16 | 0.0205207 | 3.47E-01 |
| July | 0.199283 | < 2e-16 | 0.2623221 | < 2e-16 | 0.0278059 | 0.197918 |
| August | 0.213071 | < 2e-16 | 0.280068 | < 2e-16 | 0.0429622 | 0.044504 |
| September | 0.129367 | < 2e-16 | 0.187755 | < 2e-16 | -0.0215281 | 0.317831 |
| October | 0.137625 | < 2e-16 | 0.1852354 | < 2e-16 | 0.0159108 | 0.44691 |
| November | 0.053731 | 0.000258 | 0.0740877 | 9.13E-06 | -0.0064653 | 0.755711 |
| December (Base) | - | - | - | - | - | - |
| 2007 | 0.295123 | < 2e-16 | 0.2308055 | 4.13E-13 | 0.4325048 | < 2e-16 |
| 2008 | 0.253518 | < 2e-16 | 0.2036978 | 2.49E-10 | 0.3635064 | < 2e-16 |
| 2009 | 0.302015 | < 2e-16 | 0.251928 | 3E-14 | 0.4119713 | < 2e-16 |
| 2010 | 0.117637 | 0.000148 | 0.0589867 | 0.0784 | 0.2453161 | 1.34E-09 |
| 2011 | 0.056828 | 0.054696 | 0.0173833 | 0.59022 | 0.1658723 | 0.0000248 |
| 2012 | 0.041563 | 0.115151 | 0.0046466 | 0.87264 | 0.1441031 | 0.0000514 |
| 2013 | -0.028609 | 2.07E-01 | -0.0660946 | 0.00868 | 0.0866101 | 5.71E-03 |
| 2014 | -0.103697 | 7.18E-08 | -0.1439114 | 2.5E-11 | 0.008296 | 7.59E-01 |
| 2015 | -0.025528 | 0.044677 | -0.010616 | 0.45516 | -0.0091624 | 0.616307 |
| 2016 (Base) | - | - | - | - | - | - |

Table 2 shows that the estimates for adjusted retail gasoline price are not statistically significant in any of the three models, suggesting that the fluctuation of gasoline price does not have a strong correlation with the total fatal crash, as well as male and female fatal crash numbers. This finding is in-line with the finding of the Alabama study which found that there was no significant difference between male and female drivers in response to the gasoline price change (*10*). However, this finding opposes to the finding of a previous study (*9*), which found that increasing the gasoline price more strongly reduced crashes involving female drivers than male drivers. One reason for this discrepancy could be the focus of the mentioned study on a single state. Implementation of speed limits with a maximum of 65 MPH has a significant effect in reducing female fatal crashes. Moreover, the implementation status of the differential speed limit policy for each state, which means to have different speed limits for cars and trucks, was studied. While some previous studies on differential speed limits showed adverse safety effects, some others found positive or negligible impacts (*33, 34, 35*). Till 2007, 10 states (Arkansas, California, Idaho, Illinois, Indiana, Michigan, Montana, Oregon, Texas, and Washington) were implementing the differential speed limit policy; in 2008, Illinois stopped implementing the rule, and in 2011, Texas dropped out of the list, leaving the other eight states to keep the rule till 2017. In the results, it is found that states with differential speed limit have an average of around 5% less number of fatal crashes. One possible reason for this is that large trucks take longer to decelerate or stop, making them riskier in emergent situations if they are allowed to operate at higher speed. Additionally, it seems that male and female drivers react differently to the differential speed limit policy. Compared to the states without the differential speed limit, the states with differential speed limit experience 4.4% less male fatal crashes and 8.1% less female fatal crashes. As expected, the number of fatal crashes is strongly correlated with the VMT. Higher VMT shows that more crashes are likely to occur. The results indicate that a 50% increase in VMT causes about 42.91% more fatal crashes each month. Regarding the seasonally adjusted unemployment rate, the model indicates that with a one-percent increase in the state unemployment rate, the number of fatal crashes is decreased by 3.8%. A similar conclusion was made in a study from Michigan conducted by Wagenaar, in which he claimed that the net effect of the higher unemployment rate is associated with a decline in crash involvement (*36*). The higher number of fatal crashes might be caused by the chain effects of unemployment. When more people are unemployed, the work-related travel will be reduced significantly, so they have less exposure to the risk of traffic accidents. Furthermore, the time trend variables in the models illustrate that summer months typically have more fatal crashes. Additionally, although the seat belt use rate increased by 6% from 2009 to 2016 (*37*), the number of fatal crashes increased by 2.89 % during the period; also the results do not show any correlation between the seatbelt usage and male, female and total fatal crash numbers; so, it shows that numerous other factors play roles in increasing the number of fatal crashes. Percentage of male among the population showed a significant relationship with the total fatal crash numbers; as this number goes up the total number of fatal crashes increases. Population density represented to have a negatively significant relationship with the number of female crashes. This means areas that as the population density increases the number of female fatal crashes increases as well.

Individual models are developed for motorcycles, non-motorcycles, pedestrians and bicycles to investigate the effects of gasoline price changes on the number of fatal crashes.

**TABLE 3 Results of Random Effect Negative Binomial Regression Models for Fatal Crash Numbers (Motorcycle and Non-motorcycle) at the Monthly Level in the US, 2007-2016**

| Random Effects: Group Name, State (Intercept), Number of obs: 6120, groups:State:51 | Motorcycle | | Non-motorcycle | | Pedestrian | | Bicyclist | |
|---|---|---|---|---|---|---|---|---|
| | Variance | Std. Dev. | Variance | Std. Dev. | Variance | Std. Dev. | Variance | Std. Dev. |
| | 0.0844 | 0.2905 | 0.07043 | 0.2654 | 0.1178 | 0.3432 | 0.2191 | 0.4681 |
| Fixed Effects: | Estimate | Pr(>\|z\|) | Estimate | Pr(>\|z\|) | Estimate | Pr(>\|z\|) | Estimate | Pr(>\|z\|) |
| (Intercept) | -11.438655 | 0.0428 | -4.737946 | 0.045547 | -5.6742278 | 4.30E-01 | -2.475446 | 0.758945 |
| Ln(VMT/Population) | 1.029756 | < 2e-16 | 0.918114 | < 2e-16 | -0.1610141 | 0.009449 | -0.194368 | 0.017745 |
| Adjusted Retail Gasoline Price (2018 Dollars) | 0.218222 | 3.74E-14 | -0.018728 | 0.015869 | -0.0840291 | 0.082899 | 0.097078 | 0.164567 |
| 65-mph Speed Limit and below | -0.039767 | 0.60971 | 0.024508 | 0.375049 | -0.5157876 | 2.95E-05 | -0.591714 | 0.000492 |
| 70-mph Speed Limit | -0.025394 | 0.70724 | 0.040762 | 0.090402 | 0.269268 | 0.017965 | 0.258432 | 0.090758 |
| 75-mph Speed Limit and Above (Base) | - | - | - | - | - | - | - | - |
| Differential Speed Limit | -0.01218 | 0.84737 | -0.065747 | 0.000294 | 0.0248186 | 0.846454 | 0.01755 | 0.920354 |
| Seatbelt Usage Rate | - | - | -0.179323 | 0.096961 | - | - | - | - |
| West Coast | -0.051605 | 0.79421 | -0.561044 | 0.00065 | -0.229322 | 0.379017 | -0.23736 | 0.474786 |
| East Coast | -0.227932 | 0.15838 | -0.377997 | 0.014086 | 0.0841381 | 0.700367 | -0.004268 | 0.988031 |
| Rocky Mountain | -0.148725 | 0.45359 | -0.485181 | 0.007138 | -0.4862016 | 0.066329 | -0.379178 | 0.269696 |
| Midwest | -0.336081 | 0.02727 | -0.304851 | 0.028673 | -0.1701929 | 0.405109 | -0.177852 | 0.496553 |
| Gulf Coast (Base) | - | - | - | - | - | - | - | - |
| Seasonally Adjusted Unemployment Rate (%) | -0.028655 | 0.00586 | -0.036828 | < 2e-16 | -0.0007718 | 0.967857 | 0.015344 | 0.577748 |
| Percent of 70-years-old and above | 0.058286 | 0.07124 | -0.010465 | 0.54814 | -0.1206116 | 0.009589 | -0.055399 | 0.326842 |
| Male Percentage | 0.078267 | 0.46361 | 0.035709 | 0.42298 | 0.2530001 | 0.062045 | 0.129855 | 0.395895 |
| Ln (Population Density) | 0.002157 | 0.96843 | -0.111495 | 0.009185 | 0.1480935 | 0.04031 | 0.107294 | 0.231767 |
| January | -0.003779 | 0.94939 | -0.098 | 8.40E-13 | -0.2398726 | 0.00353 | -0.076253 | 0.54526 |
| February | 0.080183 | 0.17278 | -0.191851 | < 2e-16 | -0.369504 | 7.66E-06 | -0.146637 | 0.248552 |
| March | 0.799534 | < 2e-16 | -0.094236 | 1.26E-11 | -0.3556373 | 2.14E-05 | 0.01866 | 0.883029 |
| April | 1.199551 | < 2e-16 | -0.085277 | 1.65E-09 | -0.4354903 | 3.32E-07 | 0.217762 | 0.086113 |
| May | 1.525334 | < 2e-16 | -0.030945 | 0.030688 | -0.397264 | 5.13E-06 | 0.399653 | 0.001819 |
| June | 1.661764 | < 2e-16 | -0.025941 | 0.070709 | -0.4747611 | 6.33E-08 | 0.50414 | 8.01E-05 |
| July | 1.710773 | < 2e-16 | 0.00655 | 0.643037 | -0.3678831 | 2.13E-05 | 0.62572 | 6.06E-07 |
| August | 1.761792 | < 2e-16 | 0.012939 | 0.356044 | -0.3180322 | 0.000209 | 0.586767 | 2.66E-06 |
| September | 1.540186 | < 2e-16 | -0.047838 | 0.000705 | -0.2128966 | 0.012403 | 0.597242 | 1.55E-06 |
| October | 1.20328 | < 2e-16 | 0.028022 | 0.039637 | -0.0339476 | 0.68265 | 0.423299 | 0.000556 |
| November | 0.630266 | < 2e-16 | 0.006778 | 0.615469 | -0.0483869 | 0.555738 | 0.273482 | 0.025817 |
| December (Base) | - | - | - | - | - | - | - | - |
| 2007 | -1.31735 | < 2e-16 | 0.272788 | < 2e-16 | 0.2993417 | 0.015512 | -0.40359 | 0.016637 |
| 2008 | -1.373304 | < 2e-16 | 0.212885 | 5.38E-14 | 0.3748897 | 2.95E-03 | -0.406305 | 0.020281 |
| 2009 | -1.140424 | < 2e-16 | 0.23577 | 5.22E-16 | 0.3368272 | 9.36E-03 | -0.461701 | 0.011739 |
| 2010 | -1.235531 | < 2e-16 | 0.216503 | 1.36E-13 | 0.3498726 | 0.008114 | -0.414574 | 0.026953 |
| 2011 | -1.429225 | < 2e-16 | 0.173472 | 4.98E-10 | 0.4068575 | 0.003042 | -0.471002 | 0.015908 |
| 2012 | -1.360686 | < 2e-16 | 0.151801 | 9.42E-10 | 0.4249707 | 0.001073 | -0.484371 | 0.008954 |
| 2013 | -1.481655 | < 2e-16 | 0.097866 | 4.26E-06 | 0.3847459 | 0.001171 | -0.396735 | 0.018629 |
| 2014 | -1.513182 | < 2e-16 | 0.029247 | 1.03E-01 | 0.3949741 | 0.000199 | -0.429841 | 0.00485 |
| 2015 | -0.071187 | 0.0581 | -0.01109 | 0.349006 | 0.0281137 | 7.03E-01 | 0.023632 | 0.820288 |
| 2016 (Base) | - | - | - | - | - | - | - | - |

According to Table 3, similar to the results of the previous three models, both motorcycle and non-motorcycle fatal crashes are strongly and positively correlated with the VMT. Population is found to be negatively correlated with pedestrian and bicycle fatal crashes, however by looking at the population density trend that incorporates the effect of area size, it is found that this variable is significantly correlated with the number of pedestrian fatal crashes with a positive relationship, which means as the population density increases, the number of pedestrian fatal crashes increases as well. Population density is negatively correlated with the non-motorcycle fatal crashes, which means as the population density goes higher, the number of non-motorcycle fatal crashes decreases. The probable reason for this observation is that areas with lower population density could be mostly located in rural regions. Several studies in the literature has shown that fatal crash density in rural areas is significantly higher than the same value in the urban areas (*38*); probable reasons for this observation include more speeding, dangerous roads and not wearing seatbelt in rural areas (*39*). The gasoline price changes are strongly correlated with motorcycle and non-motorcycle fatal crashes. The motorcycle model estimates suggest that one dollar increase in the retail gasoline price, which is the primary variable of interest in this study, leads to 24.2% more motorcycle fatal crashes; it means that when gasoline price changes, people tend to change their mode of transportation. For example, when the gasoline price is higher, individuals are more likely to drive a motorcycle to save fuels. The increasing inclination toward the motorcycle could contribute to the higher number of motorcycle fatal crashes. On the other hand, one dollar increase in the retail gasoline price results in a 1.9% decrease in non-motorcycle fatal crashes; the negative relationship between number of non-motorcycle fatal crashes and retail gasoline price could be the consequence of the fact that in reaction to the increasing gasoline price, people drive less and they drive in a less aggressive manner. The gasoline price changes is slightly correlated with the pedestrian fatal crashes with a negative relation. The underlying reason might be due to the people's transportation vehicle change as described above. There is less probability for crash occurrence between motorcycles and pedestrians than non-motorcycles and pedestrians due to the smaller common colliding section on their surface, which might lead to fewer fatal crashes for pedestrians as the gasoline price increases. The gasoline price changes is not correlated with the bicyclist fatal crashes; it might had been expected to observe a similar shift to using bicycles and walking at the time of increased gasoline price at a less extent but according to (*7*), there are several reasons that motorcycle crashes are a more serious problem than crashes involving bicyclists and pedestrians: 1- Motorcyclist are riding in a much higher speed than bicyclists or pedestrians and consequently the released energy in motorcycle crashes is significantly greater; 2- Motorcyclists do not travel on walkways or special lanes that separate them from the vehicular traffic and the probability of hitting a much bigger vehicle is higher; 3- Motorcycles are more feasible for traveling long distances and during inclement weather conditions, thus higher gasoline price are expected to affect motorcyclist more than bicyclists or pedestrians. The state maximum speed limit does not seem to have significant effects on the overall motorcycle and non-motorcycle state fatalities, however it plays an important role in pedestrian and bicyclist fatal crash numbers and it is strongly correlated with the number of fatal crashes among those groups; the speed limit of 65 miles per hour and below has a negative coefficient in both the bicyclist and pedestrian fatal crash trends, which shows that in comparison to the speed limit of 75 MPH and higher, it has significantly lower number of fatal crashes. The impacts of maximum speed limit might be more apparent if the regression models are developed for different road types. The differential speed limit of cars and trucks has positive effects on the non-motorcycle fatal crash numbers. It is observed that the number of non-motorcycle fatal crashes are 6.5% lower in the states with differential speed limit policy. Nevertheless, this policy does not help to decrease the number of motorcycle fatal crashes. This value was not statistically significant for pedestrian and bicyclist fatal crashes as they do not travel on highways where this value comes into play. Similar to the previous model results, seasonally adjusted unemployment rate correlates to the number of motorcycle and non-motorcycle fatal crashes. With a one percent growth in the state unemployment rate, the number of motorcycle and non-motorcycle fatal crashes declines 2.9% and 3.7% respectively. The negative relationship between the number of fatal crashes and unemployment rates is observed. The probable reason is that people drive less and reduce their daily commutes or other unnecessary trips when they do not have proper employment status. Higher drops are observed for non-motorcycle fatal crashes, which could be explained by the fact that non-motorcycle drivers are in the majority and if someone owns both motorcycle and non-motorcycle vehicles, due to the economic considerations, he or she might use the motorcycle more often for the daily trips. Regarding regional districts, results demonstrate that the West Coast, Rocky Mountain, and East Coast states have less non-motorcycle fatal crashes compared to other districts respectively. Percent of 70-years-old and above has a positive correlation with motorcycle fatal crashes, which shows that older motorcyclists have higher number of fatal crashes, but this trend is the opposite for pedestrians, as the percent of 70-years-old people rises, the pedestrian fatal crash number decreases. Moreover, the time trend variables in the model point out that the number of fatal crashes peaks in the summer months, especially for motorcycle and bicycle fatal crashes. One explanation for this observation is that the individuals are more inclined to travel or drive faster during the summer. Also, motorcycles are less likely to be used in inclement weather conditions.

## CONCLUSION

In conclusion, this study examines the relationship between gasoline prices and the number of fatal crashes. The study suggests that higher gasoline price leads to less non-motorcycle and pedestrian fatal crashes and more motorcycle fatal crashes; a one dollar increase in the gasoline price reduces the number of fatal crashes by 1.9% and increases the number of motorcycle fatal crashes by 24.2%. Overall, male and female drivers do not react differently to the gasoline price changes. The findings provide policymakers a vision of how traffic safety is going to respond to the gasoline price changes. It should be noted that higher gasoline prices help reduce non-motorcycle and pedestrian fatal crashes while sharply increase the motorcycle fatal crashes. Merely increasing the retail gasoline price or imposing higher gasoline taxes will not directly improve the traffic safety. The consequences of increasing gasoline prices should be further evaluated to improve the overall roadway safety. Raising the gasoline price might have other potential benefits. For instance, people might drive less, hence, cutting down the gasoline demand and consumption, and consequently, benefitting the environment. Also, less congestion might be expected as a result of less travel. Despite all the advantages that the increasing gasoline price might bring to the society, it still requires more serious and thorough investigations, mainly finding a way to enhance the motorcycle safety. As mentioned previously, by raising the gasoline price, more motorcycle fatal crashes would occur, and all drivers have to spend more money on the transportation, and this may make people cancel some valuable trips. Decision makers should find a balance between the cost and benefits of changing the gasoline price and adjust the policies accordingly.

One of the limitations of this study is that due to the lack of the data, instead of using the average monthly gasoline price for each state, the average monthly gasoline price for each district was used. The second limitation is that the study only focuses on the fatal crash data. A more comprehensive study could be conducted if the data for less severe crashes were included in the analysis. For the future study, it is recommended to include the weather and road type data and also the data regarding the helmet use among the motorcyclists as it can improve the fatal crash prediction models.